\documentclass[12pt]{article}
\usepackage{multicol}
\usepackage{amssymb}
\usepackage{subfig}
\usepackage{color}
\usepackage{epsfig,amssymb,amsfonts,amsmath,graphicx,dsfont,cite,xfrac}
\usepackage{authblk}
\usepackage{tikz-cd}

\definecolor{mygray}{gray}{0.5}

\title{Non-Hermitian propagation in equally-spaced Hermitian waveguide arrays}

\author[1,*]{Ivan Bocanegra}
\author[1]{H\'ector M. Moya-Cessa}

\affil[1]{Instituto Nacional de Astrof\'isica \'Optica y Electr\'onica, Calle Luis Enrique Erro No.1, Santa Mar\'ia Tonanzintla, Puebla, 72840, M\'exico}
\affil[*]{Corresponding author: ibocanegra@inaoep.mx }

\date{}
\begin{document}

\maketitle

\begin{abstract}
A non-unitary transformation leading to a Hatano-Nelson problem is performed on an array of equally-spaced optical waveguides. Such transformation produces a non-reciprocal system of waveguides, as the corresponding Hamiltonian becomes non-Hermitian. This may be achieved by judiciously choosing an attenuation (amplification) of the injected (or exciting) field. The non-Hermitian transport induced by such transformation is studied for several cases and closed analytical solutions are obtained. The corresponding non-Hermitian Hamiltonian may represent an open system that interacts with the environment, either loosing to or being provided with energy from the exterior. 

\end{abstract}

\section{Introduction}

Waveguide arrays or optical lattices are indeed pretty interesting optical devices for many reasons. In the last decades they have attracted considerable attention as they present novel diffraction phenomena, compared with the corresponding phenomena appearing in the bulk. Examples in linear as well as non-linear optics \cite{Christodoulides2003} can be readily mentioned. In such optical devices the discrete diffraction can be controlled by manipulating the properties of the array \cite{Morandotti1999,Eisenberg2000,Pertsch2002}. Also, waveguide arrays have served to study and simulate a big number of both classical and quantum  phenomena like quantum walks \cite{Perets2008}, coherent and squeezed states \cite{Perez-Leija2010,Leon-Montiel2010,Leon-Montiel2011,Villegas-Martinez2022}, Talbot effect \cite{Iwanow2005,Rai2008}, among others \cite{Szameit2007,Bromberg2009,Rodriguez-Lara2011}, in both the relativistic and non-relativistic \cite{Longhi2010,Dreisow2009} regimes. Furthermore, they have direct applications in the management of optical information. The most popular example at hand is probably the directional coupler \cite{Yariv}. In this context it is quite exceptional the implementation of waveguide arrays as mode converters described in Ref. \cite{Heinrich2014}.

A particular well-studied example is the equally-spaced waveguide array. It is defined by a uniform distance
between each pair of adjacent waveguides (upper panel in Figure \ref{fig.coupled}) and is characterized by presenting \textit{ballistic transport} \cite{Christodoulides2003,Perets2008,Eichelkraut2013}. Three regimes can be distinguished: a) the array has a semi-infinite \cite{Leon-Montiel2010} number $n$ of waveguides (it extends to $n\to\infty$), b) it has a finite \cite{Pertsch2002,Makris2006,Perets2008,Soto-Eguibar2011} number $N$ of waveguides, or c) it has a completely infinite \cite{Jones1965} number $n\in\mathbb Z$ of waveguides  (lower panel in Figure \ref{fig.coupled}). Apart from equally-spaced, there exist some other well-known arrays in which the distance between adjacent waveguides is not uniform (see for instance \cite{Perez-Leija2010,Rodriguez-Lara2011}), possessing a wide range of quite interesting features in its transport too.

In recent years, optical lattices have also been studied in the non-Hermitian scheme. In this framework PT-symmetry \cite{Bender1998,Bender1999,Mostafazadeh2003}, where the system is invariant under simultaneous parity (P) and time-reversal (T) operations (also see \cite{Ruter2010}), plays a leading role, exhibiting a variety of new behaviour such as double refraction, power oscillations, phase transitions at exceptional points \cite{Makris2008,Makris2010}, invisibility \cite{Regensburger2012,Longhi2015}, Bloch oscillations \cite{Wimmer2015} and transition from ballistic to diffusive transport \cite{Eichelkraut2013}, to mention a few. 

More specifically, non-Hermitian lattices have been studied under the Hatano-Nelson model \cite{Liu2021,Liu2022,Weidemann2020}, which in turn is a simple non-unitary transformation of a standard Schr\"odinger equation
\begin{equation}
    i\frac{\partial |\Psi(t)\rangle}{\partial t}=\left[\frac{\hat{p}^2}{2}+V({x})\right]|\Psi(t)\rangle.
\label{Schr}
\end{equation}
By doing  
\begin{equation}
|\Psi(t)\rangle=e^{\alpha x} |\Phi(t)\rangle, \qquad \alpha\in\mathbb R,
\label{transformation}
\end{equation}
the non-Hermitian Schr\"odinger equation \cite{Hatano}
\begin{equation}
    i\frac{\partial |\Phi(t)\rangle}{\partial t}=\left[\frac{(\hat{p}+i\alpha)^2}{2}+V({x})\right]|\Phi(t)\rangle,
    \label{nSchr}
\end{equation}
is reached. In fact, non-unitary transformations naturally lead to non-Hermitian dynamics \cite{Braulio}. In the optical scheme, a transformation like (\ref{transformation}) leads to a non-reciprocal or anisotropic \cite{Weidemann2020} system of waveguides. This can be interpreted as an open system (interacting with the environment) since the corresponding Hamiltonian is non-Hermitian. Indeed, the relation (\ref{transformation}) can be inverted, giving the solutions of the non-Hermitian system (\ref{nSchr}) in terms of those of the Hermitian one (\ref{Schr}), through a quite simple transformation.

Then, along the following lines we study the non-Hermitian transport of an actual Hermitian equally-spaced array, by applying a transformation similar to (\ref{transformation}) on a given initial state of the Hermitian system. This leads to an \textit{effective} non-reciprocal (non-Hermitian) system of waveguides, for which the aforementioned regimes (semi-infinite, finite and infinite) are dealt with.  

The outline of the paper is as follows: in section \ref{sec.equally} some generalities about equally-spaced waveguide arrays are given. Besides, the method to obtain the non-Hermitian transport of the Hermitian system is established in this section, so we consider this section the cornerstone of the manuscript. In sections \ref{sec.semi}, \ref{sec.finite} and \ref{sec.infinite} the semi-infinite, finite and completely infinite cases are tackled, respectively, based on the founding ideas of section \ref{sec.equally}. Finally, in section \ref{sec.conclusions} the main conclusions are drawn.
\section{Equally-spaced waveguide array}
\label{sec.equally}

\begin{figure}[h]
    \centering
{\includegraphics[width=300pt]{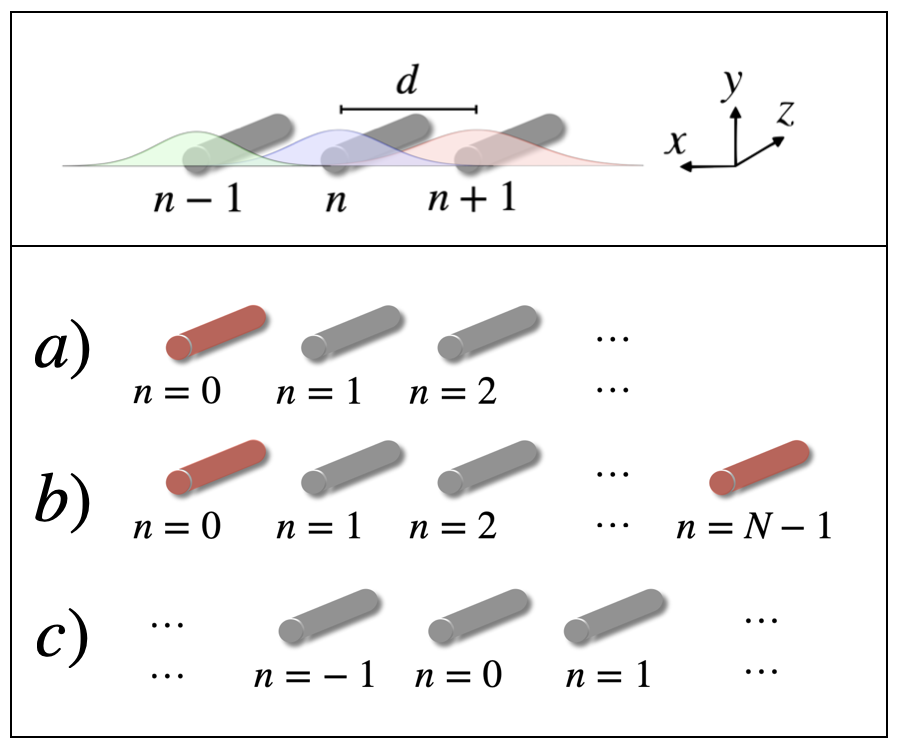}}
\caption{Upper panel: schematic coupling of electromagnetic modes between adjacent waveguides for fixed $z$. The modes are represented by Gaussian functions  of the electric field amplitude in the transversal $x$ direction and $d$ is the distance between waveguides. The coupling takes place where the Gaussian distributions overlap, for instance, by means of the evanescent fields in the space between waveguides. Lower panel: for the equally-spaced array of waveguides, it is possible to distinguish between a) semi-infinite, b) finite and c) completely infinite arrays. In a) the array begins, for instance at $n=0$, however there is no upper limit for $n$, \textit{i.e.} the array extends to $n\to\infty$. Therefore, the array has only one edge (or one end), at $n=0$. The edge waveguide is highlighted in red color. In b) a finite number $N$ of waveguides is considered, $n$ ranging from $0$ to $N-1$ for example, such that there are two edge waveguides (as well in red). In c) the array is infinite towards both $n\to\pm\infty$ such that no edges are present.}
	\label{fig.coupled}
\end{figure}

In a tight-binding optical lattice \cite{Yariv}, the evolving amplitude $c_n(z)$ of the electric field on the $n$-th waveguide,  $n\in S\subseteq \mathbb Z$, of the array is coupled to the neighbouring field amplitudes $c_{n-1}(z)$ and $c_{n+1}(z)$ of the adjacent waveguides by means of evanescent fields in the transversal direction (upper panel in Figure \ref{fig.coupled}). In the case of equally-spaced waveguides, the electric field amplitude $c_n(z)$ is given by \cite{Christodoulides2003,Makris2006}
\begin{equation}
    i\dot c_n(z)= \alpha\left[c_{n-1}(z)+c_{n+1}(z)\right],
\label{optical}
\end{equation}
where $\alpha\propto e^{-d}\in\mathbb R
$ is the coupling constant, $d$ is the distance between adjacent waveguides and the dot denotes derivative with respect to $z$. Equation (\ref{optical}) has an associated equation of the Schr\"odinger type (a partner equation \cite{Perez-Leija2010}, see also Ref. \cite{Soto-Eguibar2011}) 
\begin{equation}
    i\frac{\partial |\psi(z)\rangle}{\partial z} = H|\psi(z)\rangle,
\label{quantum}
\end{equation}
where the Hamiltonian operator
\begin{equation}
    H=\alpha(V^{{\dagger}}+V),
\label{Hamilt}
\end{equation}
for a pair of well-defined conjugate ladder operators $V$ and $V^\dagger$, is Hermitian, namely $H=H^{\dagger}$. 
An abstract Hilbert space $\mathcal H$ is assumed, where each $c_n(z)$ is associated to an abstract vector $|n\rangle\in\mathcal H$. The partner equations (\ref{optical}) and (\ref{quantum}) are connected by 
\begin{equation}
|\psi(z)\rangle=\sum_{n} c_n(z)|n\rangle,
\label{linear}
\end{equation}
where $V^\dagger|n\rangle=|n+1\rangle$ and  $V|n\rangle=|n-1\rangle$ (for simplicity the reader can think of a completely infinite array, \textit{i.e.} $n\in\mathbb Z$). The state $|\psi(z)\rangle\in\mathcal H$ represents the total electric field amplitude in the array for all $z$ and can be normalized by means of $\displaystyle\sum_{n} |c_n(z)|^2=1$, for instance.

Depending on whether the waveguide array into consideration is either finite, semi-infinite or completely infinite, there will be different starting and ending sites (lower panel in Figure \ref{fig.coupled}). The operators $V^{\dagger}$ and $V$ in (\ref{Hamilt}) as well as the electric field amplitudes 
\begin{equation}
c_{k}(z)=\langle k|\psi(z)\rangle,\qquad k\in S,
\label{ceka}
\end{equation}
will be defined accordingly. 

In order to set the frame for the generation of the non-Hermitian propagation in the equally-spaced (Hermitian) system generic ladder operators $V^{\dagger}$ and $V$ will be considered in what follows. The specific definition of such operators for each case (finite, semi-infinite and infinite) and the set of vectors $\left\{|n\rangle\right\}$ will then be given in the corresponding section.
\subsection{Non-Hermitian transport in the Hermitian equally-spaced system}

When the Hamiltonian (\ref{Hamilt}) is $z$-independent, the solution of (\ref{quantum}) is given by
\begin{equation}
    |\psi(z)\rangle=e^{-iHz}|\psi(0)\rangle,
\label{CI}
\end{equation}
with $|\psi(0)\rangle$ a given initial condition, for instance at $z=0$.

A transformation
\begin{equation}
|\psi(0)\rangle=e^{-\gamma \hat n} |\phi(0)\rangle,
\qquad \gamma\in\mathbb R,
\label{initial}
\end{equation}
is considered, with $\hat n |k\rangle=k|k\rangle$, $|k\rangle\in\mathcal H$.

By replacing (\ref{initial}) into (\ref{CI}), it is obtained
\begin{equation}
    |\psi(z)\rangle=e^{-\gamma \hat n}e^{\gamma \hat n}e^{-iHz}e^{-\gamma \hat n}|\phi(0)\rangle=e^{-\gamma \hat n}|\phi(z)\rangle,
\label{psiphi}
\end{equation}
with 
\begin{equation}
    |\phi(z)\rangle:=e^{\gamma \hat n}e^{-iHz}e^{-\gamma \hat n}|\phi(0)\rangle=e^{-i \tilde{H} z}|\phi(0)\rangle,
\label{phiz}
\end{equation} 
the solution of the non-Hermitian (non-conservative) problem
\begin{equation}
    i\frac{\partial |\phi(z)\rangle}{\partial z} = \tilde H|\phi(z)\rangle,
\label{nHermitian}
\end{equation}
where
\begin{equation}
    \tilde H=\alpha(e^{ \gamma}V^{\dagger}+e^{- \gamma}V)=k_1V^{\dagger}
+k_2V,
\label{htilde}
\end{equation}
 is non-Hermitian, this is $\tilde H^\dagger\neq \tilde H$, and $k_1=\alpha e^\gamma, k_2=\alpha e^{-\gamma}$. We have used the commutation relations $[\hat n,V]=-V$ and $[\hat n,V^\dagger]=V^\dagger$ to obtain the last expression in (\ref{phiz}). From now on we focus on the non-Hermitian problem (\ref{nHermitian})-(\ref{htilde}).
 
In the particular case of the infinite array the non-Hermitian Hamiltonian (\ref{htilde}) coincides with Eq. (2) of Ref. \cite{Longhi2015}. Thus, the problem (\ref{nHermitian}) for the non-Hermitian Hamiltonian in (\ref{htilde}), \textit{i.e.} the Hatano-Nelson problem, is rather interesting by itself. Moreover, a detailed proposal of its implementation in optical devices by means of coupled resonator optical waveguides (CROW's) is finely described in \cite{Longhi2015} (also see \cite{Liu2021,Liu2022,Weidemann2020} for alternative implementations). In a different approach, here we analyze (\ref{nHermitian})-(\ref{htilde}) as modelling the non-Hermitian transport of the actual Hermitian system (\ref{quantum})-(\ref{Hamilt}) in the finite, semi-infinite and infinite regimes. 

Equation (\ref{nHermitian}) has a partner equation
\begin{equation}
    i\dot d_n(z)= k_1 d_{n-1}(z)+k_2d_{n+1}(z),
\label{neoptical}
\end{equation}
by means of
\begin{equation}
|\phi(z)\rangle=\sum_{n} d_n(z)|n\rangle,
\label{dlinear}
\end{equation}
where 
\begin{equation}
    d_k(z)=\langle k
|\phi(z)\rangle=e^{\gamma k}c_k(z), \qquad
k\in S,
\label{dn}
\end{equation} represents the amplitude of the electric field propagating in the $k$-th waveguide of the non-conservative (non-Hermitian) array. To obtain the last equality in (\ref{dn}), (\ref{ceka}) and (\ref{psiphi}) have been used.

\begin{figure}[h]
    \centering
    {\includegraphics[width=300pt]{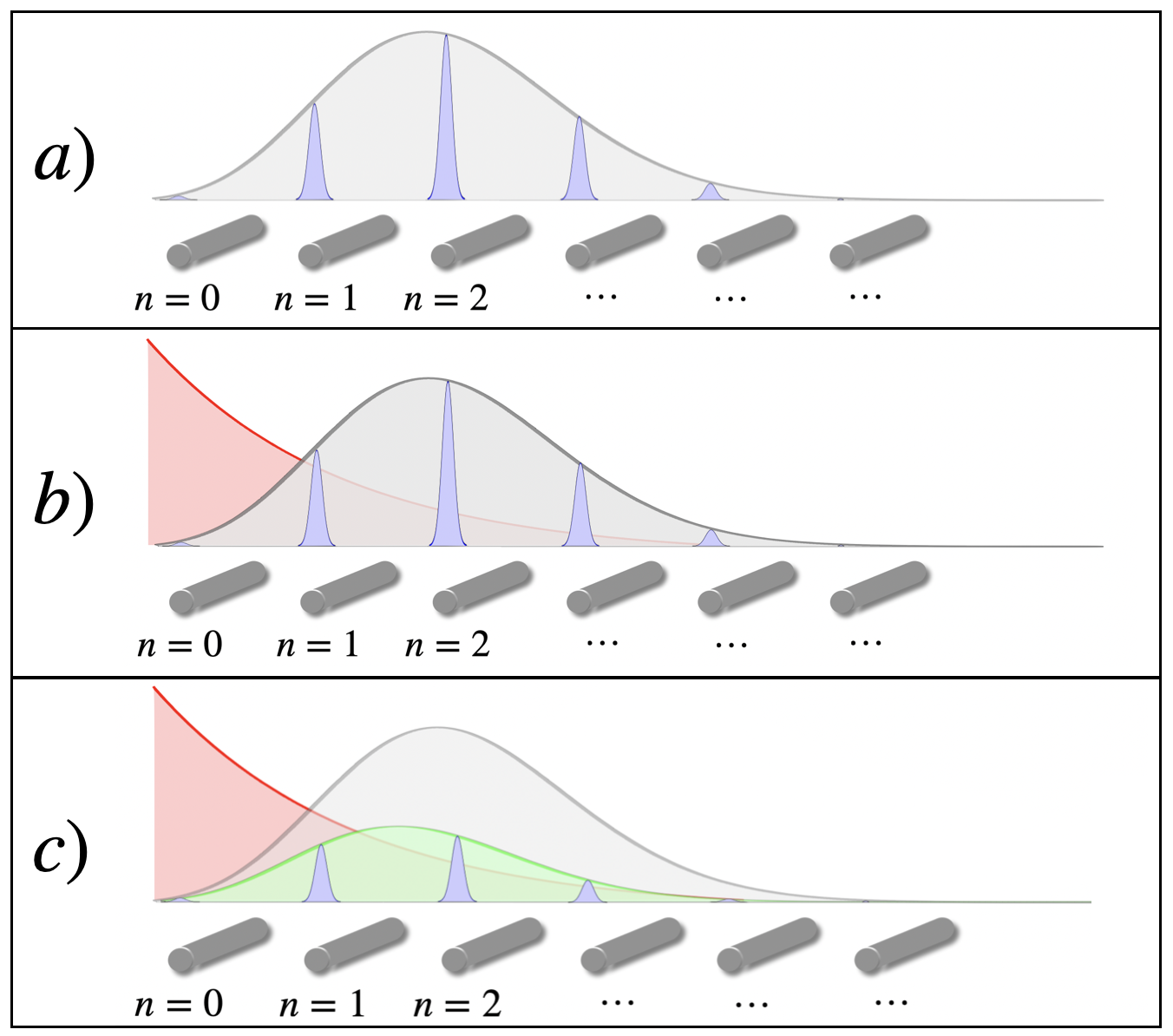}}
\caption{Schematic representation of the non-unitary transformation (\ref{phipsi}) on an initial state $|\psi(0)\rangle$ with Poissonian distribution of coefficients $c_n(0)$, for a semi-infinite waveguide array. a) The modes of each waveguide are represented by blue Gaussian functions, whose heights are defined by $c_n(0)$ for each $n$, and the envelope of the Poisson distribution is shown in gray color. The horizontal represents the transversal coordinate $x$, as in the upper panel of Figure \ref{fig.coupled}.  b) The transformation (\ref{phipsi}) is roughly represented by a red exponentially decreasing ($\gamma<0$) operation $e^{\gamma \hat n}$ on $|\psi(0)\rangle$. The resulting state $|\phi(0)\rangle=e^{\gamma \hat n}|\psi(0)\rangle$ is sketched in c). The 
envelope of the 
 resulting coefficients $d_n(0)=e^{\gamma n}c_n(0)$, according to (\ref{dn}), is shown in green color. Naturally, in the case $\gamma>0$ the transformation (\ref{phipsi}) becomes an amplification of the initial state $|\psi(0)\rangle$ increasing with $n$.}
	\label{fig.transformation}
\end{figure}

For general $k_1$ and $k_2$, the solution $|\phi(z)\rangle$ of (\ref{nHermitian})-(\ref{htilde}) can be cast in terms of the solution $|\psi(z)\rangle$ of the corresponding Hermitian problem (\ref{quantum})-(\ref{Hamilt}) as
\begin{equation}
  |\phi(z)\rangle=  \left(\frac{k_1}{k_2}\right)^{\displaystyle\frac{\hat n}{2}}|\psi(z)\rangle .
\label{gphipsi}
\end{equation}
For $k_1$ and $k_2$ as given in (\ref{htilde}), it is seen that (\ref{gphipsi}) is indeed equivalent to (\ref{psiphi}). Then, the non-Hermitian transport is obtained from the Hermitian one by means of the rather simple relation
\begin{equation}
 |\phi(z)\rangle=  e^{\gamma\hat n}|\psi(z)\rangle .
\label{phipsi}
\end{equation}
Figure \ref{fig.transformation} schematically shows the effect of the non-unitary transformation (\ref{phipsi}) on an initial state $|\psi(0)\rangle$. It represents an (external) exponential  attenuation ($\gamma<0$) or amplification ($\gamma>0$) of the initial electric field distribution. The resulting state
$|\phi(0)\rangle=e^{\gamma \hat n}|\psi(0)\rangle$ (and more generally $|\phi(z)\rangle$) is not normalized because the system is either being provided with external energy or its energy is being damped by an outer process, according to transformation (\ref{phipsi}). In turn, such attenuation/amplification can be genuinely achieved in the laboratory \cite{Ruter2010}.

\begin{figure}[h]
    \centering
{\includegraphics[width=300pt]{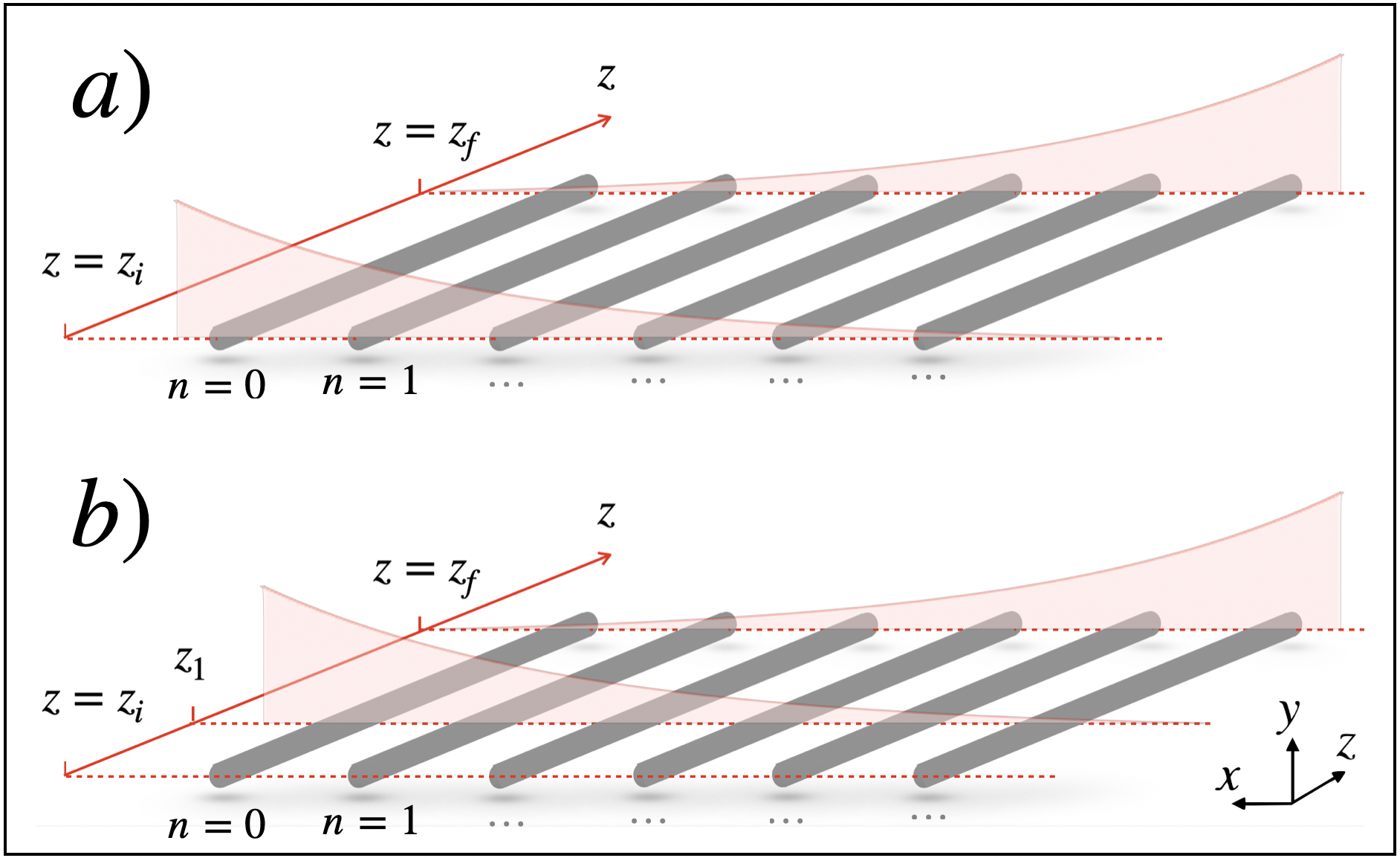}}
\caption{Representation of the non-unitary transformation (\ref{phipsi}) on the waveguide array, and its effect along the propagation in $z$. a) At $z=z_i$ the transformation (\ref{phipsi}) is performed on the initial state $|\psi(z_i)\rangle$. Later at $z=z_f$, the inverse transformation (\ref{psiphi}) is applied on $|\phi(z_f)\rangle$, taking us back to $|\psi(z_f)\rangle$. In the interval $z_i\leq z<z_f$ the state $|\phi(z)\rangle$ models the non-Hermitian transport of the actual Hermitian system. It is remarkable that the transformation (\ref{phipsi}) can be applied at any $z$, not necessarily $z=0$. In b) such transformation is schematically applied at some $z_1>z_i$ and the inverse transformation (\ref{psiphi}) is once more applied at $z_f$. Thus, for $z_1\leq z< z_f$, $|\phi(z)\rangle$ models non-Hermitian propagation of the electromagnetic field. Certainly, for $z<z_1$ and $z\geq z_f$ we have Hermitian propagation, thus one can alternate intervals of Hermitian and non-Hermitian transport at will.}
	\label{fig.transport}
\end{figure}

In turn, Figure \ref{fig.transport} shows how to apply transformation (\ref{phipsi}) in order to obtain the non-Hermitian propagation of the Hermitian system and how to recover the Hermitian solution by means of the inverse transformation (\ref{psiphi}). It is worth to mention that, as the non-unitary transformation (\ref{phipsi}) can be performed at any $z$, the same as (\ref{psiphi}), it is possible to alternate intervals of Hermitian and non-Hermitian propagation at will.
\section{Semi-infinite array}
\label{sec.semi}
In the semi-infinite case the operators $V$ and $V^\dagger$ (the Susskind-Glogower operators) appearing in the Hamiltonian in (\ref{Hamilt}) are defined (together with $\hat n$) as 
\begin{subequations}
    \begin{equation}
    V=\sum_{k=0}^{\infty} |k\rangle\langle k+1|,
    \end{equation}\begin{equation}V^{{\dagger}}=\sum_{k=0}^{\infty}|k+1\rangle\langle k|,
    \end{equation}
    \begin{equation} \hat{n}=\sum_{k=0}^{\infty}k |k\rangle\langle k|.
    \end{equation}
\end{subequations}
The corresponding commutation relations are (see \cite{Leon-Montiel2010} and references therein)
\begin{subequations}
    \begin{equation}[V,V^\dagger]=|0\rangle\langle 0|,
    \end{equation}
    \begin{equation}
    [\hat n,V]=-V,
    \end{equation}
    \begin{equation}
    [\hat n,V^\dagger]=V^\dagger.
    \end{equation}
\label{commutation}
\end{subequations}
And the relation between
(\ref{optical}) and (\ref{quantum}) is given by 
\begin{equation}
    |\psi(z)\rangle=\sum_{k=0}^\infty c_k(z)|k\rangle,
\label{combination}
\end{equation}
with $\left\{|k\rangle\right\}_{k=0,1,\dots}$ the Fock states. Due to the semi-infinite nature of the problem, the condition $c_{-j}(z)=0$, for $j=1,2,\dots$, needs to be added, so that for $n=0$, (\ref{optical}) becomes
\begin{equation}
    i\dot c_0(z)= \alpha c_{1}(z).
\end{equation}
In parallel, we have 
\begin{equation}
    i\dot d_0(z)= k_2 d_{1}(z),
\end{equation}
added to (\ref{neoptical}).

The ladder operators above may be written in terms of annihilation and creation operators as
\begin{equation}
    V=\frac{1}{\sqrt{\hat{n}+1}}a,\qquad V^{{\dagger}}=a^{{\dagger}}\frac{1}{\sqrt{\hat{n}+1}},\qquad \hat{n}=a^{{\dagger}}a .
\end{equation}
From (\ref{CI}), we have
\begin{equation}
    |\psi(z)\rangle=e^{-i\alpha z(V^\dagger+V)}|\psi(0)\rangle\equiv D_{SG}(-i\alpha z)|\psi(0)\rangle,
\label{DSG}
\end{equation}
where $D_{SG}(\xi)=e^{\xi V^\dagger-\xi^\star V}$, $\xi\in\mathbb C$, is a displacement operator (in analogy to the Glauber displacement operator \cite{Leon-Montiel2010,Gazeau2021}) in terms of $V^\dagger$ and $V$. The star denotes complex conjugation. As explained in \cite{Leon-Montiel2010}, the first expression in (\ref{commutation}) spoils any attempt of applying the Baker-Campbell-Hausdorff formula (see Ref. \cite{Perez-Leija2010,Louisell}) on $D_{SG}(\xi)$.

If the initial condition $|\psi(0)\rangle$ is chosen such that only one waveguide is excited at $z=0$ (the impulse response of the system is desired \cite{Makris2006}), for instance the $m$-th waveguide, then we have
\begin{equation}
    |\psi(0)\rangle=|m\rangle,\qquad m=0,1,\dots
\end{equation}
Indeed, it is possible to  generalize the Eq. (11) in Ref. \cite{Leon-Montiel2010} to
\begin{equation} |m,\xi\rangle_{SG}\equiv D_{SG}(\xi)|m\rangle  = \left(1-V^{2m+2}\right)e^{\xi V}e^{\xi V^{\dagger}}|m\rangle.
\label{gcs}
\end{equation}
In the context of nonlinear coherent states, (\ref{gcs}) represents a generalized nonlinear Susskind-Glogower (SG) coherent state (see \cite{Leon-Montiel2011}) and the Eq. (11) in Ref. \cite{Leon-Montiel2010} can be recovered by setting $m=0$.

\begin{figure}[h]
    \centering
    {\includegraphics[width=350pt]{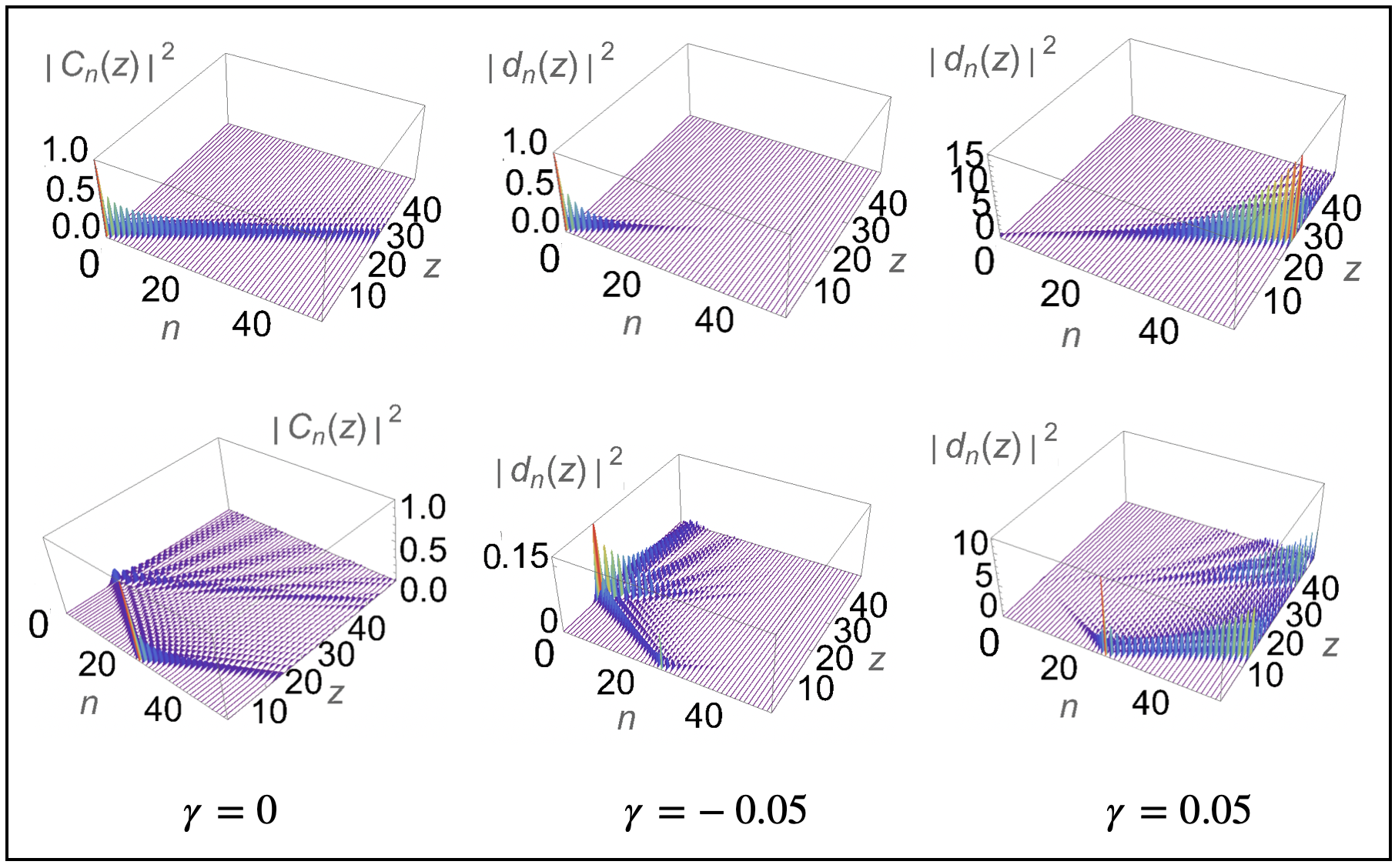}}
\caption{Comparison between the Hermitian and non-Hermitian transport of the electromagnetic field intensity $|d_n(z)|^2$, according to (\ref{dn1}), in an equally-spaced semi-infinite waveguide array when a single site $m$ is initially excited at $z=0$. The distance $z$ is measured in units of the propagation constant $\alpha$ (here $\alpha=1$ unit has been chosen). Upper row: the edge waveguide ($m=0$) is initially excited for the Hermitian $\gamma=0$ (upper left), and non-Hermitian $\gamma\neq 0$ (upper middle and upper right) cases. In the Hermitian case (upper left) the propagation is not damped or amplified. For $\gamma = -0.05$ and $\gamma = 0.05$ (upper middle and upper right, respectively) exponential damping and amplification is observed (compare the vertical scales). The system is either loosing or being provided by external energy: it behaves as an open system. Lower row: the waveguide at site $m=25$ is initially excited. Reflection takes place at the edge waveguide due to the semi-infinite nature of the array. This certainly breaks the symmetry of the propagation along the array. In the Hermitian case (lower left) we have no damping or amplification.
For $\gamma=-0.05$ (lower middle) and $\gamma=0.05$ (lower right) the effect of the transformation (\ref{phipsi}) is clear: an attenuation and amplification, respectively, of the electromagnetic field towards $n\to\infty$. In particular for $\gamma=-0.05$ (lower middle) the so-called skin effect is observed at the edge waveguide $m=0$.}
	\label{fig.semiinfinite1}
\end{figure}

As we have $\xi=-i\alpha z$, (\ref{DSG}) results into (compare to Eq. (76) in \cite{Leon-Montiel2011})
\begin{equation}
    |\psi(z)\rangle=
    \sum_{s=0}^\infty\left[(-i)^{s-m}J_{s-m}(2\alpha z)+(-i)^{s+m}J_{s+m+2}(2\alpha z)\right]|s\rangle,
\label{psizeta}
\end{equation}
with $J_k(x)$ the Bessel function \cite{Abramowitz} of order $k$. Finally, $c_n(z)=\langle n|\psi(z)\rangle$ is obtained as
\begin{equation}
    c_n(z)= (-i)^{n-m}J_{n-m}(2\alpha z)+(-i)^{n+m}J_{n+m+2}(2\alpha z),
\label{cn1}
\end{equation}
$n,m=0,1\dots$, coinciding with the Eq. (3) in \cite{Makris2006}. In turn, the electric field amplitudes in the non-conservative system are simply
\begin{equation}
    d_n(z)=e^{\gamma n}\left[(-i)^{n-m}J_{n-m}(2\alpha z)+(-i)^{n+m}J_{n+m+2}(2\alpha z)\right].
\label{dn1}
\end{equation}
For $m=0$, the previous expression reduces to (compare to \cite{Gazeau2021})
\begin{equation}
    d_n(z)=e^{\gamma n}\frac{1}{\alpha z}(-i)^{n}(n+1)J_{n+1}(2\alpha z).
\end{equation}
Figure \ref{fig.semiinfinite1} shows the electromagnetic intensities $|d_n(z)|^2$ for some values of the parameters, showing the non-Hermitian propagation dictated by (\ref{dn1}) and its comparison with the Hermitian case $\gamma=0$. When the edge waveguide $m=0$ is initially excited (upper row), we have no amplification or attenuation in the Hermitian case $\gamma=0$ (upper left). For $\gamma=-0.05$ (upper middle) the electromagnetic field suffers attenuation as it propagates. Indeed, this case describes more accurately an actual physical propagation, as losses due to the interaction with the environment are always present. On the other hand, for $\gamma=0.05$ an exponential amplification is observed. This, in turn, describes a situation in which, external electromagnetic power is provided to the system, for instance, in order to compensate for losses. On the other hand, when a waveguide in the bulk is initially excited, for instance at $m=25$ (lower row), the presence of the boundary at $m=0$ becomes evident. The reflection at the edge waveguide can be better appreciated in the Hermitian case (lower left). In turn, for the cases $\gamma=-0.05$ and $\gamma=0.05$ (lower middle and lower right, respectively) the effect of the non-unitary transformation (\ref{phipsi}) is clear, an attenuation and amplification, respectively, towards $n\to\infty$. In particular, for $\gamma=-0.05$ (lower middle) the electromagnetic field is redistributed mainly towards the boundary, at $m=0$. This effect is known in the literature as the non-Hermitian skin effect and has been both studied and implemented in a variety of different kind of systems \cite{Liu2021,Liu2022,Weidemann2020}.
\section{Finite array}
\label{sec.finite}
In the case of a finite array, the operators $V$ and $V^\dagger$ in (\ref{Hamilt}), together with $\hat n$, are defined by
\begin{subequations}
    \begin{equation}
    V_N=\sum_{k=0}^{N-1} |k\rangle\langle k+1|, 
    \end{equation}
    \begin{equation}
    V_N^{{\dagger}}=\sum_{k=0}^{N-1} |k+1\rangle\langle k|, 
   \end{equation}
    \begin{equation} \hat{n}_{N}=\sum_{k=0}^{N-1}k |k\rangle\langle k|,
    \end{equation}
\label{semioperators}
\end{subequations}
 where the subindex $N$ has been added to emphasize that they correspond to the finite case, where we have a total of $N$ waveguides in the array. The commutation relations are
\begin{subequations}
    \begin{equation}[V_N,V_N^\dagger]=|0\rangle\langle 0|,
    \end{equation}
    \begin{equation}
    [\hat n_N,V_N]=-V_N,
    \end{equation}
    \begin{equation}
    [\hat n_N,V_N^\dagger]=V_N^\dagger.
    \end{equation}
\label{commuta}
\end{subequations}
Indeed, these have exactly the same form of those in (\ref{commutation}). Thus, along this section we only call $V$, $V^\dagger$ and $\hat n$, respectively, the operators defined in (\ref{semioperators}).

We connect
(\ref{optical}) and (\ref{quantum}) through 
\begin{equation}
    |\psi(z)\rangle=\sum_{k=0}^{N-1} c_k(z)|k\rangle,
\end{equation}
with $|k\rangle\in\mathcal H, k=0,1,\dots,N-1$. As now we have two edge or ending waveguides, besides (\ref{optical}) we have 
\begin{equation}
    i\dot c_0(z)=\alpha c_1(z), \qquad i\dot c_{N-1}(z) = \alpha c_{N-2}(z),
\label{opticalfinite}
\end{equation}
and
\begin{equation}
    i\dot d_0(z)=k_2 d_1(z), \qquad i\dot d_{N-1}(z) = k_1 d_{N-2}(z),
\end{equation}
besides (\ref{neoptical}).
In order to follow \cite{Soto-Eguibar2011}, we change to a normalized variable $Z:=\alpha z$, such that (\ref{optical}) and (\ref{opticalfinite}) become
\begin{equation}
    i\dot c_n(Z) = c_{n-1}(Z)+c_{n+1}(Z),
\end{equation}
\begin{equation}
    i\dot c_0(Z) = c_{1}(Z),
\end{equation}
\begin{equation}
    i\dot c_{N-1}(Z) = c_{N-2}(Z).
\end{equation}
The Hamiltonian (\ref{Hamilt}), for $V$ and $V^\dagger$ as given in (\ref{semioperators}), then takes the simple form of a square tri-diagonal matrix of size $N$ (see also \cite{Efremidis2005})
\begin{equation}
    H= \left(\begin{array}{cccccc}
        0 & 1 & 0 & 0 & \cdots & 0\\
        1 & 0 & 1 & 0 &\cdots & 0\\
        0 & 1 & 0 & 1 &\cdots & 0\\
        \vdots & \vdots & \vdots & \ddots & & 
        \vdots\\
          0 & 0 & 0 & \cdots & 0 & 1\\
        0 & 0 & 0 & \cdots & 1& 0\\
    \end{array}\right).
    \label{Hfinite}
\end{equation}
 A similarity transformation \begin{equation}
 \Lambda = S^{-1}HS,
 \label{similarity}
 \end{equation}
 is to be performed, with $S$ an orthonormal matrix whose columns are the normalized eigenvectors $|\psi_j\rangle$ of $H$, this is
 \begin{equation}
      H |\psi_j\rangle=\lambda_j |\psi_j\rangle, \qquad j=0, 1, \dots,N-1.
\label{eva}
 \end{equation}

The $N$ eigenvalues $\lambda_0, \lambda_1,\dots,\lambda_{N-1}$ are as usual obtained from the condition
$D_N = 0$,
with $D_N\equiv$ det$(H-\lambda \mathbb I)$ and $\mathbb I$ the identity matrix of size $N$. In turn, $D_N$ can be obtained by the three-terms recurrence relation
\begin{subequations}
    \begin{equation}
  D_0=1,  
    \end{equation}
    \begin{equation}
    D_1= \lambda, 
    \end{equation}
    \begin{equation}
    D_s=\lambda D_{s-1}-D_{s-2}, \qquad s=2,3,\dots,N,
    \end{equation}
\label{recurrence}
\end{subequations}
and the corresponding normalized eigenvectors turn out to be
\begin{equation}
    |\psi_j\rangle=\left(\sum_{k=0}^{N-1} \left[D_k(\lambda_j)\right]^2\right)^{-1/2}\left(\begin{array}{c}
         D_0(\lambda_j) \\
         D_1(\lambda_j)\\
         \vdots\\ 
         D_{N-1}(\lambda_j)
          
    \end{array}\right).
\end{equation}
The recurrence relations (\ref{recurrence}) are those of the Chebyshev polynomials \cite{Abramowitz} of the second kind, such that the condition to find out the eigenvalues $\lambda_j$ of $H$ is given by 
\begin{equation}
    D_N(\lambda)=U_N\left(\frac{\lambda}{2}\right)=0,
\end{equation}
with $U_k(x)$ the Chebyshev polynomial of the second kind of order $k$. Thus, obtaining (for details, the reader is referred to \cite{Soto-Eguibar2011})
\begin{equation}
    \lambda_j= 2\cos\left[\frac{\left(j+1\right)\pi}{N+1}\right],\qquad j=0, 1, \dots,N-1.
\end{equation}
Note that unlike Ref. \cite{Soto-Eguibar2011} we label the eigenvalues from $0$ to $N-1$.
The eigenvectors thus become
\begin{equation}
   |\psi_j\rangle =\left(\sum_{k=0}^{N-1} \left[U_k(\lambda_j/2)\right]^2\right)^{-1/2}\left(\begin{array}{c}
         U_0(\lambda_j/2) \\
         U_1(\lambda_j/2)\\
         \vdots\\ 
         U_{N-1}(\lambda_j/2)
          
    \end{array}\right).
\end{equation}

By inverting (\ref{similarity}) and replacing into (\ref{CI}) we obtain
\begin{equation}
|\psi(Z)\rangle=Se^{-i\Lambda Z}S^{-1}|\psi(0)\rangle\equiv R|\psi(0)\rangle,
\label{RR}
\end{equation}
where $\Lambda$ is a diagonal matrix with diagonal elements $\lambda_0, \lambda_1,\dots,\lambda_{N-1}$, and $R$ is a square matrix of size $N$ with elements
\begin{equation}
    R_{p,q}(Z)=\sum_{j=0}^{N-1}\frac{U_{p-1}(\lambda_j/2)U_{q-1}(\lambda_j/2)\exp(-iZ\lambda_j) }{\sum_{k=0}^{N-1}\left[U_k(\lambda_j/2)\right]^2},
\label{Rpq}
\end{equation}
with $p,q=1,2,\dots,N$. From (\ref{RR}) and (\ref{Rpq}) we obtain
\begin{equation}
    c_n(Z)=\sum_{\ell=0}^{N-1}R_{n+1,\ell+1}(Z) c_\ell(0), \qquad n=0,1,\dots N-1,
\end{equation}
or explicitly
\begin{equation}\nonumber
    c_n(Z)=\sum_{\ell=0}^{N-1}\sum_{j=0}^{N-1}\frac{c_\ell(0) \exp\left(-2iZ\cos\left[\frac{\left(j+1\right)\pi}{N+1}\right]\right)}{\displaystyle\sum_{k=0}^{N-1}\left[U_k\left(\cos\left[\displaystyle\frac{\left(j+1\right)\pi}{N+1}\right]\right)\right]^2}
\end{equation}
\begin{equation}
 \times\left\{U_n\left(\cos\left[\frac{\left(j+1\right)\pi}{N+1}\right]\right)U_\ell\left(\cos\left[\displaystyle\frac{\left(j+1\right)\pi}{N+1}\right]\right)\right\} .
\end{equation}

\begin{figure}[h]
    \centering
    {\includegraphics[width=350pt]{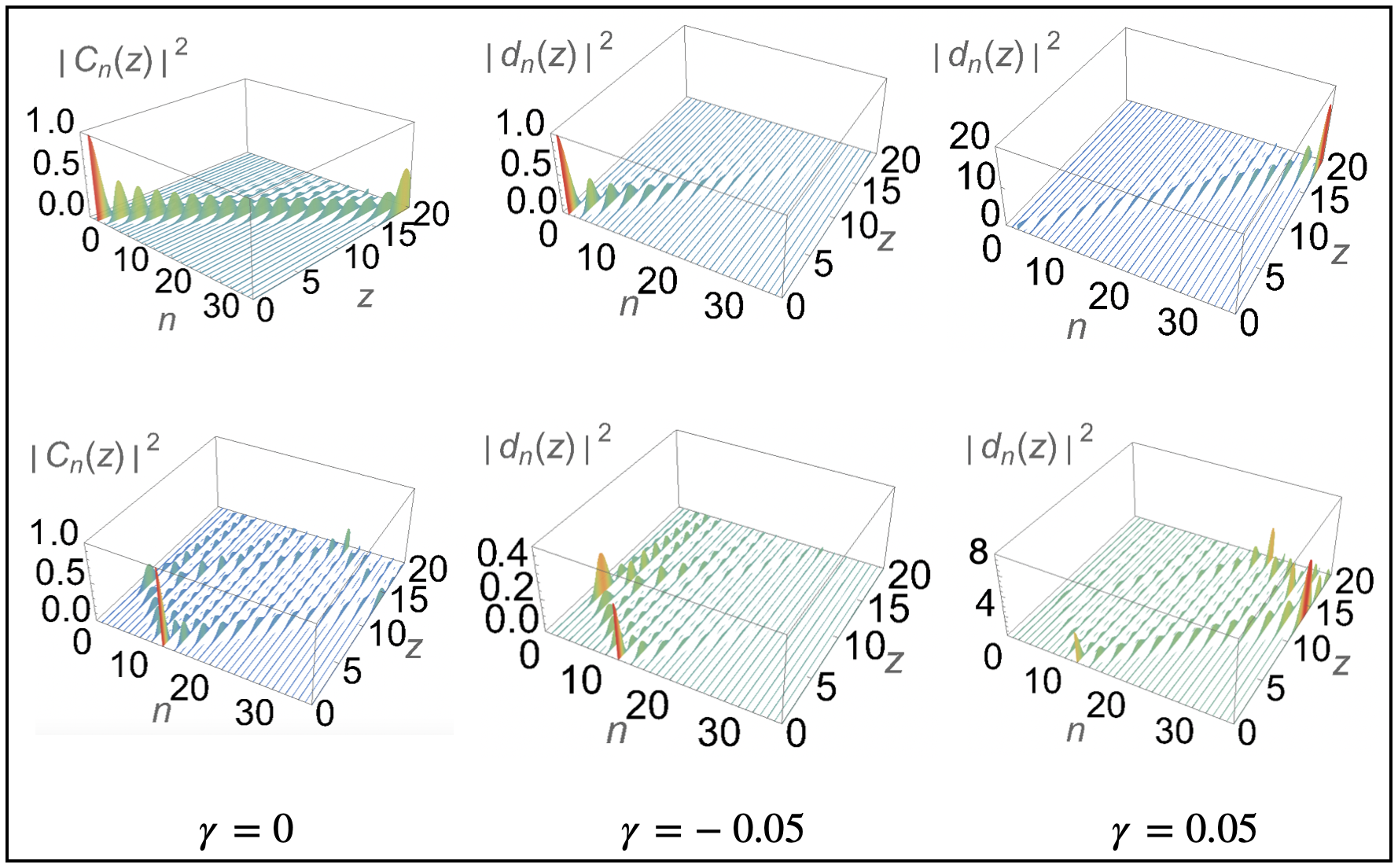}}
\caption{Comparison between the Hermitian and non-Hermitian transport of the electromagnetic field intensity $|d_n(z)|^2$, as given by (\ref{finited}), in an equally-spaced finite waveguide array consisting of $N=36$ waveguides when a single site $m$ is initially excited at $z=0$. The distance $z$ in the figure is actually $Z$ in (\ref{finited}). Upper row: the edge waveguide $m=0$ is initially excited. Similar to the upper row in Figure \ref{fig.semiinfinite1}, the Hermitian $\gamma=0$ (upper left) and non-Hermitian cases $\gamma=-0.05$ (upper middle) and  $\gamma=0.05$ (upper right) are characterized, respectively, by no attenuation or amplification, attenuated and amplificated transport (see the vertical scales) corresponding to closed (upper left) and open systems (upper middle and upper right). Due to the finite nature of the array, the larger attenuation (amplification) for $\gamma=-0.05$ ($\gamma=0.05$) is given at the edge $n=35$ in the non-Hermitian cases (upper middle and right, respectively). Lower row: the site $m=10$ is initially excited. For the Hermitian case $\gamma=0$ (lower left) reflections at both edges of the array can be clearly appreciated, in contrast to the single-side reflection in the semi-infinite case (Figure \ref{fig.semiinfinite1}). In turn, for the non-Hermitian cases $\gamma=-0.05$ (lower middle) and $\gamma=0.05$ (lower right), the left and right reflections, respectively, eclipse the corresponding reflection at the opposite edge of the array, showing again the non-Hermitian skin effect due to the non-unitary transformation (\ref{phipsi}).}
	\label{fig.finite1}
\end{figure}

For a single input as initial state $|\psi(0)\rangle$, at the $m$-th waveguide for instance, $c_\ell(0)=\delta_{\ell m}$, with $\delta_{ij}$ the Kronecker delta, thus giving
\begin{equation}\nonumber
    c_n(Z)=\sum_{j=0}^{N-1}\frac{\exp\left(-2iZ\cos\left[\frac{\left(j+1\right)\pi}{N+1}\right]\right)}{\displaystyle\sum_{k=0}^{N-1}\left[U_k\left(\cos\left[\displaystyle\frac{\left(j+1\right)\pi}{N+1}\right]\right)\right]^2}
\end{equation}
\begin{equation}
   \times \left\{U_n\left(\cos\left[\frac{\left(j+1\right)\pi}{N+1}\right]\right)U_m\left(\cos\left[\frac{\left(j+1\right)\pi}{N+1}\right]\right) \right\}.
\label{finitec}
\end{equation}
Therefore, the impulse response of the non-conservative system is simply 
\begin{equation}\nonumber
    d_n(Z)=e^{\gamma n}\sum_{j=0}^{N-1}\frac{\exp\left(-2iZ\cos\left[\frac{\left(j+1\right)\pi}{N+1}\right]\right)}{\displaystyle\sum_{k=0}^{N-1}\left[U_k\left(\cos\left[\displaystyle\frac{\left(j+1\right)\pi}{N+1}\right]\right)\right]^2}
\end{equation}
\begin{equation}
   \times \left\{U_n\left(\cos\left[\frac{\left(j+1\right)\pi}{N+1}\right]\right)U_m\left(\cos\left[\frac{\left(j+1\right)\pi}{N+1}\right]\right)\right\}.
\label{finited}
\end{equation}
Figure \ref{fig.finite1} shows the electromagnetic intensities $|d_n(z)|^2$ for some values of the parameters, according to (\ref{finited}). The distance $z$ in the Figure is actually $Z$ in (\ref{finited}). In contrast to the semi-infinite case, here we have two edges on which reflections appear (compare lower rows in Figure \ref{fig.semiinfinite1} and Figure \ref{fig.finite1}). As shown in Figure \ref{fig.finite1} (lower middle and lower right panels) such reflections can be tailored by adjusting the value of the non-Hermitian parameter $\gamma$. Again, the skin effect can be appreciated at the left and right edges in the lower middle and lower right panels of Figure \ref{fig.finite1}, respectively.
\subsection{Supermodes of the finite array}
If $|\psi(0)\rangle $ in (\ref{CI}) is chosen as $|\psi(0)\rangle =|\psi_j\rangle$, with $|\psi_j\rangle$ fulfilling equation (\ref{eva}), the (normalized) stationary solution
\begin{equation}
   |\psi(Z)\rangle=\exp\left(-i\lambda_j Z\right)|\psi_j\rangle, \qquad j=0,1,\dots,N-1,
\label{super}
\end{equation} is obtained. The solutions of the form (\ref{super}) are the supermodes of the waveguide array \cite{Yariv,Soto-Eguibar2011}.

In turn, if $|\phi(0)\rangle$ in (\ref{phiz}) is chosen as $|\phi(0)\rangle= e^{\gamma \hat n} |\psi_j\rangle$, one gets that
\begin{equation}
   |\phi(Z)\rangle=\exp\left(-i\lambda_j Z\right)e^{\gamma \hat n}|\psi_j\rangle, 
\label{super2}
\end{equation}
is stationary as well. Nevertheless, (\ref{super2}) is not normalized. 

By substitution of (\ref{super2}) into (\ref{nHermitian}), the corresponding eigenvalue equation for $\tilde H$ is reached
\begin{equation}
    \tilde H |\phi_j\rangle=\lambda_j |\phi_j\rangle,
\end{equation}
with $|\phi_j\rangle=e^{\gamma \hat n}|\psi_j\rangle$.

\begin{figure}[h]
    \centering
    {\includegraphics[width=350pt]{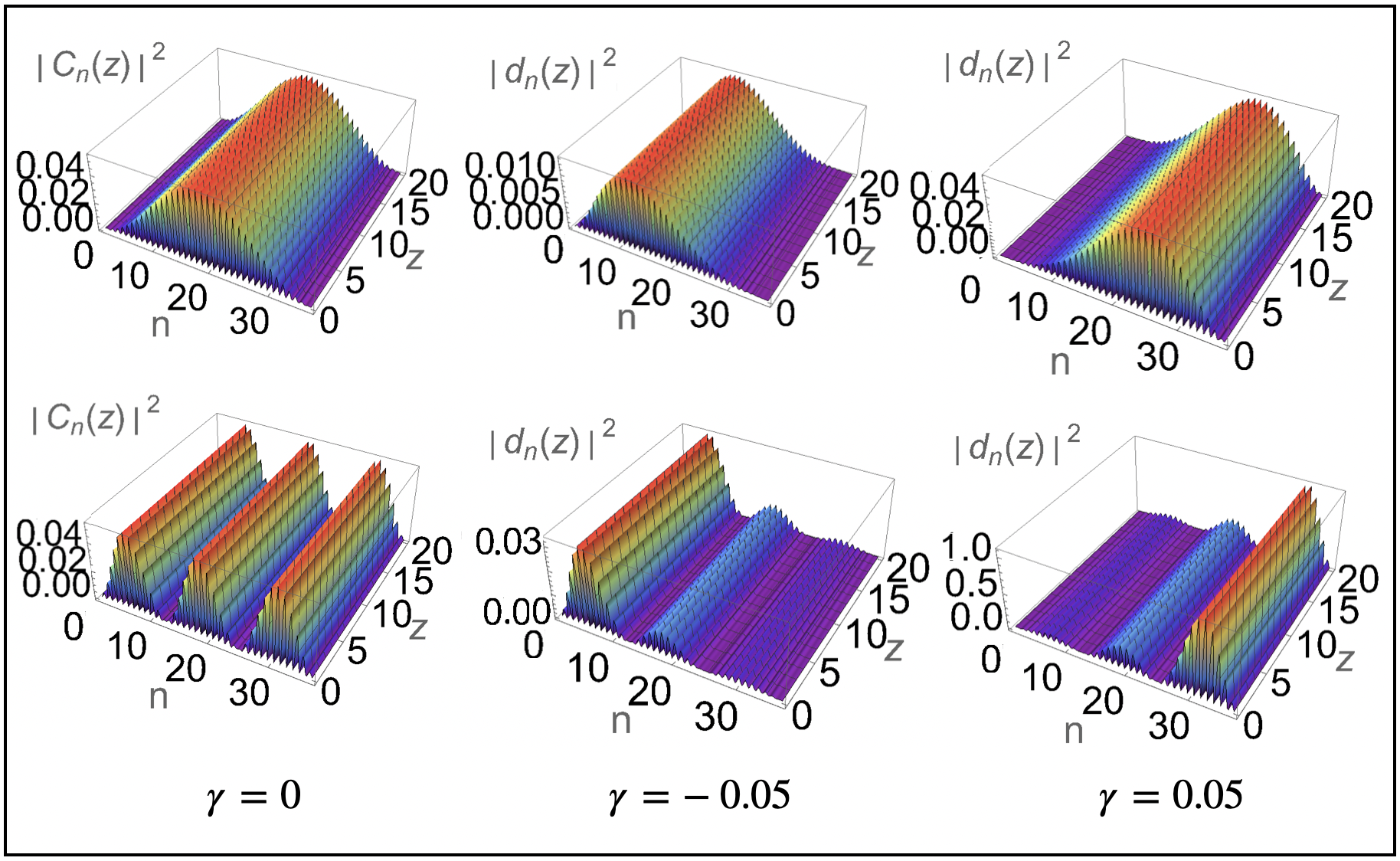}}
\caption{Numerical comparison between the eigenstates $|\psi_j\rangle$ and $|\phi_j\rangle$ of an equally-spaced finite array consisting of $N=36$ waveguides. Upper row: the first supermode ($j=0$) is sketched. The Hermitian case $\gamma=0$ (upper left) shows a symmetric distribution of the electromagnetic field centered around the middle of the array $n\sim 18$. The non-Hermitian cases $\gamma=-0.05$ (upper middle) and  $\gamma=0.05$ (upper right) are characterized by a redistribution of the electromagnetic field towards the left and right edge waveguides, with the corresponding gain or loss (see the vertical scales). Lower row: the fifth supermode $j=4$ is shown. As before, the Hermitian case $\gamma=0$ (lower left) shows a symmetric distribution of the electromagnetic field around $n\sim 18$. For the non-Hermitian cases $\gamma=-0.05$ (lower middle) and $\gamma=0.05$ (lower right), a redistribution of the electromagnetic field towards the left and right, respectively, is observed, according to the non-unitary transformation (\ref{phipsi}).}
	\label{fig.finite2}
\end{figure}

Figure \ref{fig.finite2} shows a numerical comparison between the eigenstates $|\psi_j\rangle$ (left) and $|\phi_j\rangle$ (middle and right panels), for the first $j=0$ (upper arrow) and fifth $j=4$ (lower arrow) supermodes of an equally-spaced array formed with $N=36$ waveguides. It is seen that both $|\psi(Z)\rangle$ and $|\phi(Z)\rangle$ are indeed constant along $Z$ ($z$ in the Figure). The effect roughly illustrated in Figure \ref{fig.transformation} can be clearly appreciated, \textit{i.e.} the transformation $d_n(0)=e^{\gamma n}c_n(0)$ of the initial coefficients $c_n(0)$. In this case, due to the symmetric nature of the initial distribution $c_n(0)$ around $n\sim 18$ (upper left and lower left panels), the transformation (\ref{phipsi}) causes a redistribution of the electromagnetic field maxima towards the edge waveguides. Compare with Figure 1d) in Ref. \cite{Liu2022}.

Regarding Figure \ref{fig.transport}b), one can think that we have $\gamma=0$ for $z_i\leq z<z_1$, thus obtaining the Hermitian transport displayed for instance in the upper left panel of Figure \ref{fig.finite2}. At $z=z_1$ we set $\gamma=-0.05$, getting then the non-Hermitian propagation shown in the upper middle panel of the same Figure. Afterwards, at $z=z_f$ for instance, the transformation $|\psi_0\rangle=e^{-\gamma \hat n}|\phi_0\rangle$ is performed, taking us back to the Hermitian transport shown in the upper left panel.
\section{Infinite array}
\label{sec.infinite}
In the infinite case a slight modification of equation (\ref{optical}) is considered:
\begin{equation}
    i\dot c_n(z)=\omega (-1)^nc_n(z)+ \alpha\left[c_{n+1}(z)+c_{n-1}(z)\right], \quad \omega\in\mathbb R,
    \label{infinitecase}
\end{equation}
with $n=\dots ,-2,-1,0,1,2,\dots$ The equation (\ref{infinitecase}) rules the propagation of the electromagnetic field in a binary waveguide that, in turn, represents the optical analogue of the rapid trembling motion (\textit{Zitterbewegung}) of free relativistic (Dirac) electrons, caused by the interference between positive and negative energy states in relativistic quantum mechanics, see for instance \cite{Longhi2010}. Also, a minor modification of (\ref{infinitecase}) serves to simulate Bloch-Zener oscillations in lattices of curved waveguides \cite{Dreisow2009}. The corresponding Hamiltonian is then 
\begin{equation}
H = \omega(-1)^{\hat n} +\alpha(V^\dagger+V),
\label{infiniteH}
\end{equation}
with the operators in (\ref{infiniteH}) defined by
\begin{subequations}
    \begin{equation}
    V_\infty=\sum_{k=-\infty}^{\infty} |k\rangle\langle k+1|,
    \end{equation}
    \begin{equation} V_\infty^{{\dagger}}=\sum_{k=-\infty}^{\infty} |k+1\rangle\langle k|, 
     \end{equation}
    \begin{equation} \hat{n}_{\infty}=\sum_{k=-\infty}^{\infty}k |k\rangle\langle k|.
    \end{equation}
\label{infop}
\end{subequations}
The subindex $\infty$ has been added to remark that they correspond to the infinite case. Their commutation relations are thus
\begin{subequations}
    \begin{equation}[V_\infty,V_\infty^\dagger]=[V_\infty^\dagger,V_\infty]=0,
    \end{equation}
    \begin{equation}
    [\hat n_\infty,V_\infty]=-V_\infty,
    \end{equation}
    \begin{equation}
    [\hat n_\infty,V_\infty^\dagger]=V_\infty^\dagger.
    \end{equation}
\label{infcommute}
\end{subequations}
Note that again the second and third expressions in (\ref{infcommute}) preserve the form of the corresponding expressions in (\ref{commutation}). Nonetheless, in this case $V_\infty$ and $V^\dagger_\infty$ commute. As previously, along this section the subindices are dropped by keeping in mind the definitions given in (\ref{infop}).

The relation between
(\ref{infinitecase}) and (\ref{quantum}), with $H$ in (\ref{infiniteH}), is given by
\begin{equation}
    |\psi(z)\rangle=\sum_{k=-\infty}^\infty c_k(z)|k\rangle,
\end{equation}
with $|k\rangle, k=\dots,-1,0,1,\dots$, a basis of $\mathcal H$. Namely,
\begin{equation}
    \langle m|n\rangle=\delta_{mn},\qquad \sum_{m=-\infty}^\infty|m\rangle\langle m|= \mathbb I,
\end{equation}
with $\mathbb I$ the identity operator and $m,n=\dots,-2,-1,0,1,2,\dots$ By defining $\cos{\hat{\Phi}}=\displaystyle\frac{V+V^\dagger}{2}$, the Hamiltonian (\ref{infiniteH}) becomes $H=\omega \left(-1\right)^{\hat n}+2\alpha\cos\hat\Phi$. By using the relation 
\begin{equation}
    (-1)^{\hat{n}} V = -V(-1)^{\hat{n}}, 
\end{equation}
it is obtained
\begin{equation}
  H^2= \omega^2+4\alpha^2{\cos^2{\hat{\Phi}}}.
\end{equation}
Therefore (compare to \cite{Soto-Eguibar2012})
\begin{equation}
  H^{2k}=\left(\sqrt{\omega^2+4\alpha^2{\cos^2{\hat{\Phi}}}}\right)^{2k},
\end{equation}
and
\begin{equation}\nonumber
  H^{2k+1}=\frac{1}{\sqrt{\omega^2+4\alpha^2{\cos^2{\hat{\Phi}}}}}
\end{equation}
\begin{equation}
\times\left(\sqrt{\omega^2+4\alpha^2{\cos^2{\hat{\Phi}}}}\right)^{2k+1}\left(\omega(-1)^{\hat{n}}+2\alpha{\cos{\hat{\Phi}}}\right).
\end{equation}
The propagator in (\ref{CI}) can thus be written as
\begin{equation}\nonumber
 e^{-iHz}=\cos\left(z\sqrt{\omega^2+4\alpha^2{\cos^2{\hat{\Phi}}}}\right) -i\sin\left(z\sqrt{\omega^2+4\alpha^2{\cos^2{\hat{\Phi}}}}\right)
\end{equation}
 \begin{equation}
\times\frac{1}{\sqrt{\omega^2+4\alpha^2{\cos^2{\hat{\Phi}}}}}\left(\omega(-1)^{\hat{n}}+2\alpha{\cos{\hat{\Phi}}}\right).
\end{equation}
Also, as we desire to obtain the impulse response of the system we choose $|\psi(0)\rangle=|m\rangle$. Nevertheless, due to the symmetry of the problem, we can set $m=0$ without any loss of generality. Thus, from (\ref{CI})
\begin{equation}
|\psi(z)\rangle=e^{-iHz}|0\rangle= \int_0^{2\pi}d\eta e^{-iHz}|\eta\rangle \langle\eta |0\rangle,
\label{psiz}
\end{equation}
where
\begin{equation}
    \mathbb I=\int_0^{2\pi}d\eta |\eta\rangle \langle\eta|,
\end{equation}
with $|\eta\rangle$ the phase states (see Ref. \cite{Soto-Eguibar2012}) defined as
\begin{equation}
    |\eta\rangle=\frac{1}{\sqrt{2\pi}}\sum_{n=-\infty}^{\infty}e^{in\eta}|n\rangle.
\end{equation}
By using \begin{equation}
    (-1)^{\hat n}|\eta\rangle=|\eta+\pi\rangle,\quad V^{{\dagger}}|\eta\rangle=e^{i\eta}|\eta\rangle, \quad V|\eta\rangle=e^{-i\eta}|\eta\rangle ,
\end{equation}
expression (\ref{psiz}) results into
\begin{equation}\nonumber
|\psi(z)\rangle= \frac{1}{\sqrt{2\pi}}\int_0^{2\pi}d\eta \left\{ \cos(\xi(\eta)z)|\eta\rangle\right.
\end{equation}
\begin{equation}
\left. -i\Omega(\eta,z)(2\alpha\cos\eta|\eta\rangle+\omega|\eta+\pi\rangle)\right\},
\end{equation}
with
\begin{equation}
  \Omega(\eta,z):=\frac{\sin(\xi(\eta)z)}{\xi(\eta)},\qquad \xi(\eta):=\sqrt{\omega^2+4\alpha^2\cos^2\eta}.  
\end{equation}
Also, since
\begin{equation}
    \langle k|\eta\rangle= \frac{1}{\sqrt{2\pi}}e^{ik\eta}, \qquad \langle k|\eta+\pi\rangle= \frac{1}{\sqrt{2\pi}}e^{ik\eta}(-1)^k,
\end{equation}
we obtain $c_n(z)=\langle n|\psi(z)\rangle$ as
\begin{equation}\nonumber
c_n(z)= \frac{1}{\pi}\int_0^{\pi}d\eta \left\{\cos(\eta n)\left[\cos(\xi(\eta)z)\right.\right.
\end{equation}
\begin{equation}
\left.\left. -i\Omega(\eta,z)(2\alpha\cos\eta+\omega(-1)^n)\right]\right\}.
\end{equation}
Therefore
\begin{equation}\nonumber
d_n(z)= \frac{e^{\gamma n}}{\pi}\int_0^{\pi}d\eta \left\{\cos(\eta n)\left[\cos(\xi(\eta)z)\right.\right.
\end{equation}
\begin{equation}
\left.\left. -i\Omega(\eta,z)(2\alpha\cos\eta+\omega(-1)^n)\right]\right\}.
\label{dinfinite}
\end{equation}
In the particular case $\omega=0$ in (\ref{infiniteH}), due to the first expression in (\ref{infcommute}) we have $V^\dagger=V^{-1}$. Then, by taking again $|\psi(0)\rangle=|0\rangle$, the state (\ref{CI}) becomes
\begin{equation}
    |\psi(z)\rangle=\exp\left[-\alpha z(iV-\frac{1}{iV})\right]|0\rangle=\sum_{k=-\infty}^\infty i^k J_k(-2\alpha z)V^k|0\rangle.
    \label{Jon}
\end{equation}
As $V^k|0\rangle=|-k\rangle$, the electric field amplitude $c_n(z)$ is given by
\begin{equation}
  c_n(z)=i^n J_n(-2\alpha z),
\label{Jones}
\end{equation}
in agreement with Jones \cite{Jones1965}.

\begin{figure}[h]
    \centering
   {\includegraphics[width=350pt]{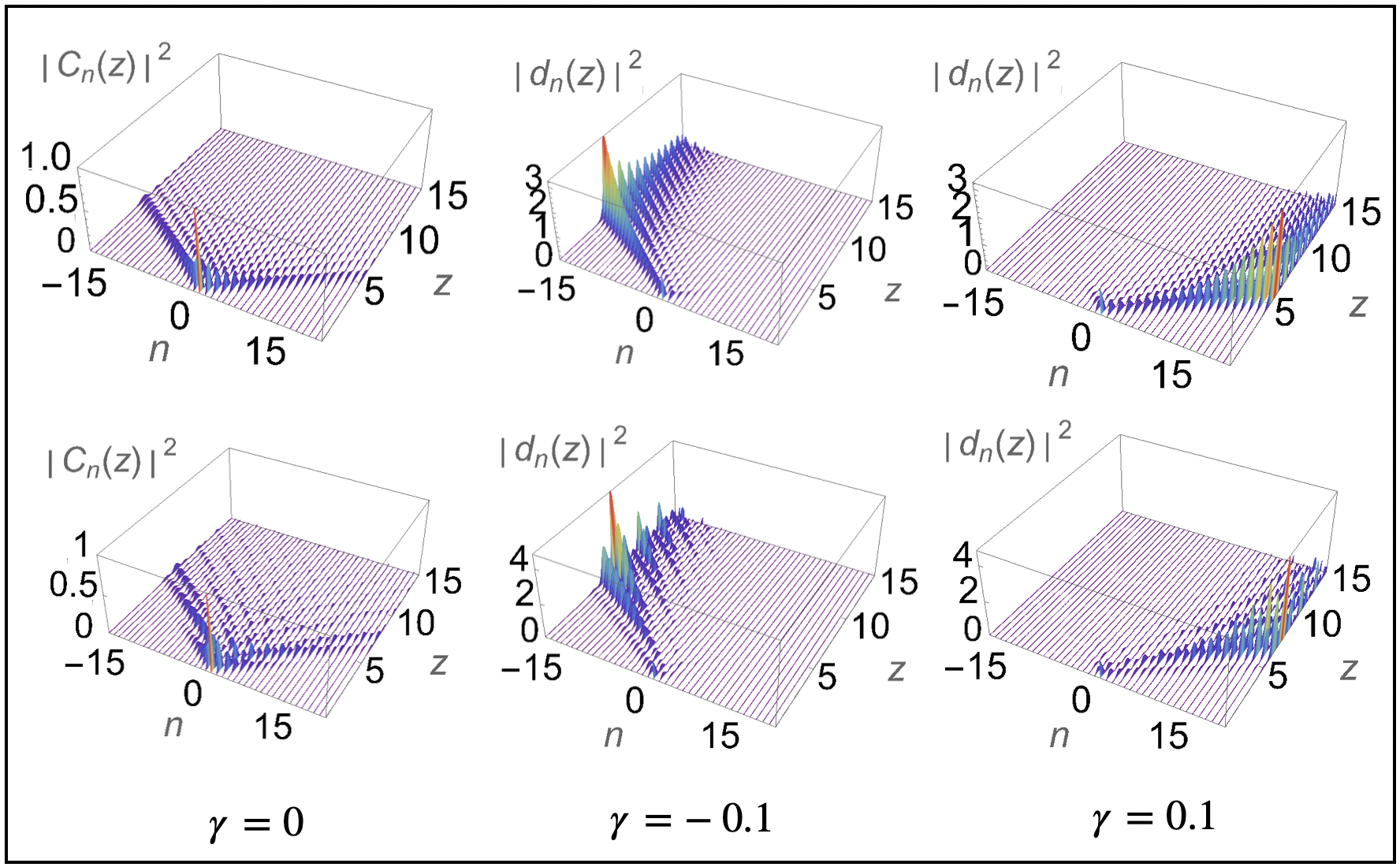}}
\caption{Comparison between the Hermitian and non-Hermitian transport of the electromagnetic field intensity $|d_n(z)|^2$, as given by (\ref{dinfinite}), in an equally-spaced infinite waveguide array when a single site $m=0$ is initially excited at $z=0$. The distance $z$ is measured in units of the propagation constant $\alpha$ (here $\alpha=2$ units has been chosen). Upper row: the simpler case $\omega=0$ is sketched. In the Hermitian case $\gamma=0$ (upper left) a well-known result of ballistic propagation is shown \cite{Christodoulides2003,Perets2008,Pertsch2002}. Unlike the previous cases (Figure \ref{fig.semiinfinite1} and Figure \ref{fig.finite1}), the inherent symmetry corresponding to the completely infinite waveguide array is preserved, as no (reflections on) edge waveguides  are present. In turn, for the non-Hermitian cases $\gamma=-0.1$ (upper middle) and $\gamma=0.1$ (upper right), the propagation to the left and right directions, respectively, is promoted (amplified) while inhibited (attenuated) towards the opposite direction. It is remarkable that, in this case, the upper middle and upper right panels are specular reflections of each other with respect to $n=0$. Lower row: the modification $\omega=1$ is shown. In the Hermitian case $\gamma=0$ (lower left) the symmetry due to the infinite nature of the array is still present, nevertheless showing a richer dynamics compared with the $\omega=0$ situation (upper left), due to the $\omega\neq 0$. In the non-Hermitian cases $\gamma=-0.1$ (lower middle) and $\gamma=0.1$ (lower right), as before, the propagation towards the left and right directions, respectively, is stimulated while suppressed towards the opposite direction, as dictated by (\ref{phipsi}). Again, the lower middle and lower right panels are specular reflections of each other with respect to $n=0$.}
	\label{fig.infinite1}
\end{figure}

Figure \ref{fig.infinite1} shows the electromagnetic field intensities $|d_n(z)|^2$, according to (\ref{dinfinite}), for some values of the parameters. The Hermitian case $\gamma=\omega=0$ (upper left panel in Figure \ref{fig.infinite1}) is well-studied \cite{Pertsch2002}. It presents perfect ballistic propagation \cite{Christodoulides2003,Perets2008,Eichelkraut2013} due to the symmetric nature of the infinite waveguide array. For the corresponding non-Hermitian transport $\gamma\neq 0$, propagation towards either increasing or decreasing $n$ can be promoted (amplified) while inhibited (attenuated) in the opposite direction (upper middle and right panels). On the other hand, the Hermitian ($\gamma=0$) case $\omega\neq 0$ (lower left) shows a more interesting dynamics compared to the corresponding $\omega=0$ choice (upper left). As before, in the corresponding non-Hermitian transport $\gamma\neq 0$ propagation towards the right or the left of the array can be enhanced while diminished in the contrary direction (lower middle and lower right panels). Unlike Figures \ref{fig.semiinfinite1} and \ref{fig.finite1}, upper (lower) middle and upper (lower) right panels in Figure \ref{fig.infinite1} are specular reflections of each other with respect to $n=0$.
\subsection{Circular waveguide array}
\begin{figure}[h]
    \centering
   {\includegraphics[width=300pt]{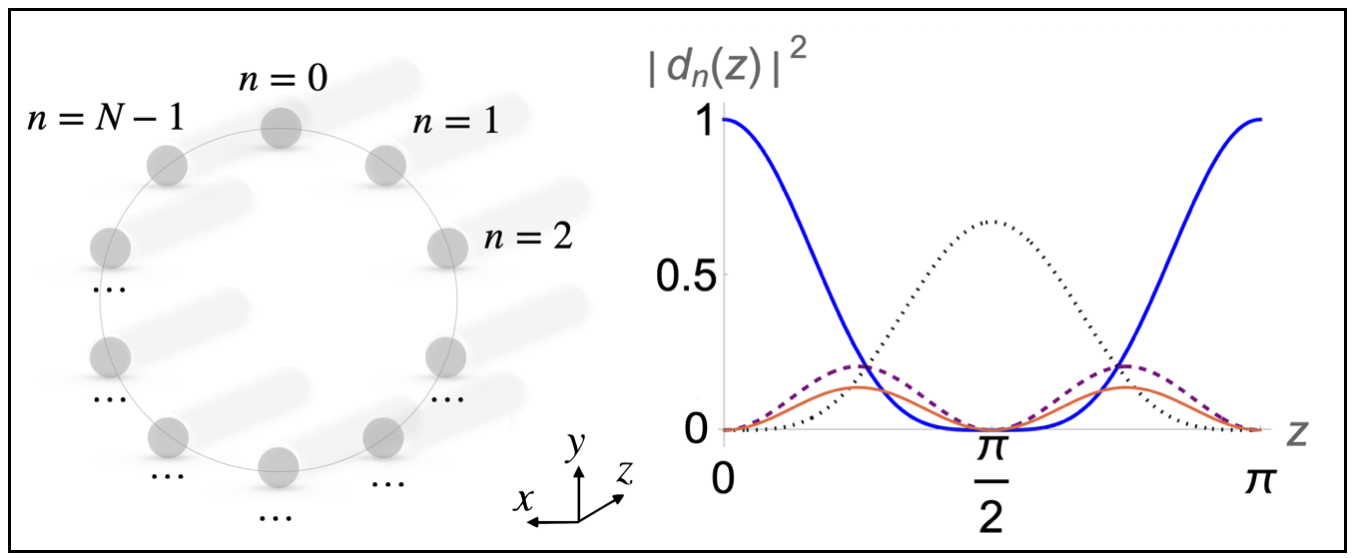}}
\caption{Left: circular array of $N$ waveguides. The electric field in the $n$-th waveguide is coupled with the
corresponding fields of the adjacent waveguides according to equations (\ref{uno})-(\ref{tres}). Right: square modulus of the fields (\ref{circ}), for $\gamma=-0.1$: $d_0(z)$ (solid blue), $d_1(z)$ (dashed purple), $d_2(z)$ (dotted black) and $d_3(z)$ (solid red). The fields are periodic with period $\pi$ and, due to the transformation (\ref{phipsi}), $|d_1(z)|^2\neq|d_3(z)|^2$, in contrast to the Hermitian ($\gamma=0$) case.}
	\label{fig.circular}
\end{figure}

There exists a finite array whose solution coincides with (\ref{Jon}). Consider a circular array of $N$ waveguides as the one shown in Figure \ref{fig.circular} (left). The electric field in the $n$-th waveguide is described by ($\alpha=1$)
\begin{equation}
    i\dot c_0(z) = c_{N-1}+c_{1}(z),
    \label{uno}
\end{equation}
\begin{equation}
    i\dot c_{N-1}(z) = c_{N-2}(z)+c_{0}(z),
\end{equation}
\begin{equation}
    i\dot c_n(z) = c_{n-1}(z)+c_{n+1}(z),\qquad n=1,2,\dots, N-2.
    \label{tres}
\end{equation}
The Hamiltonian in (\ref{Hamilt}), for instance for $N=4$, is given by
\begin{equation}
    H= \left(\begin{array}{cccc}
        0 & 1 & 0 & 1\\
        1 & 0 & 1 & 0\\
        0 & 1 & 0 & 1\\
        1 & 0 & 1 & 0\\
    \end{array}\right).
\end{equation}Note the non-zero entries in the upper-right and lower-left corners of the array, in comparison with (\ref{Hfinite}) and due to the circular nature of the array. In turn, $H=V+V^\dagger$, with \cite{Perez-Leija2016}
\begin{equation}
    V= \left(\begin{array}{cccc}
        0 & 1 & 0 & 0\\
        0 & 0 & 1 & 0\\
        0 & 0 & 0 & 1\\
        1 & 0 & 0 & 0\\
    \end{array}\right),
\end{equation}
and $V^\dagger$ the transpose of $V$. It can be readily proved that $V V^\dagger = V^\dagger V = \mathbb I$. Therefore the solution (\ref{CI}), for $|\psi(0)\rangle=|0\rangle$, coincides with (\ref{Jon}). Nevertheless, as rather generally $V^N=(V^\dagger)^N=\mathbb I$, and due to $V^{-1}=V^\dagger$, the solution (\ref{Jon}) can be written as
\begin{equation}
    |\psi(z)\rangle=\left(\sum_{s=0}^{N-1}G_s(z)(V^\dagger)^{s+1}+\sum_{s=0}^{N-1}F_s(z)V^s\right)|0\rangle,
\end{equation}
where
\begin{equation}
G_s(z)\equiv\sum_{\ell=0}^{\infty}i^{N\ell+s+1} J_{N\ell+s+1}(-2z),
\end{equation}
\begin{equation}
F_s(z)\equiv\sum_{\ell=0}^{\infty}i^{N\ell+s} J_{N\ell+s}(-2z).
\end{equation}
For $N=4$,
\begin{subequations}
\begin{equation}
c_0(z)=G_3(z)+F_0(z),
\end{equation}
\begin{equation}
c_1(z)=G_0(z)+F_3(z),
\end{equation}
\begin{equation}
c_2(z)=G_1(z)+F_2(z),
\end{equation}
\begin{equation}
c_3(z)=G_2(z)+F_1(z).
\end{equation}
\end{subequations}
The electric fields corresponding to the anisotropic system are simply
\begin{subequations}
\begin{equation}
d_0(z)=c_0(z),
\end{equation}
\begin{equation}
d_1(z)=e^\gamma(G_0(z)+F_3(z)),
\end{equation}
\begin{equation}
d_2(z)=e^{2\gamma}(G_1(z)+F_2(z)),
\end{equation}
\begin{equation}
d_3(z)=e^{3\gamma}(G_2(z)+F_1(z)).
\end{equation}
\label{circ}
\end{subequations}
Figure \ref{fig.circular} (right) shows the square modulus of fields (\ref{circ}) for specific values of the parameters. The fields are seen to be periodic with period $\pi$. Unlike the Hermitian case ($\gamma=0$), where we have $|c_1(z)|^2=|c_3(z)|^2$, in the non-Hermitian one $|d_1(z)|^2\neq|d_3(z)|^2$, due to transformation (\ref{phipsi}) (compare to Reference \cite{Perez-Leija2016}).
\section{Conclusions}
\label{sec.conclusions}
The non-Hermitian propagation induced by a non-unitary transformation, leading to a Hatano-Nelson type problem of an actual equally-spaced Hermitian waveguide array, was discussed. Three regimes were addressed: the finite, semi-infinite and completely infinite waveguide arrays. Quite generally, such transformation enhances the propagation towards one (transversal) direction in the array while inhibiting propagation towards the opposite direction. 

In the semi-infinite and finite regimes, when one edge waveguide is initially excited, the intensity $|d_n(z)|^2$ of the electromagnetic field is damped ($\gamma< 0$) as it propagates (upper middle panels in Figure \ref{fig.semiinfinite1} and Figure \ref{fig.finite1}), thus describing a more accurate situation of actual  propagation in comparison with the Hermitian $\gamma=0$ case (upper left panels in the same Figures). On the other hand, the non-Hermitian $\gamma>0$ case models external providing of power, shown for instance in the upper right panels of Figure \ref{fig.semiinfinite1} and Figure \ref{fig.finite1}. Thus, the (open) non-Hermitian system $\gamma\neq 0$ describes better the actual physical propagation phenomena in the finite and semi-infinite arrays when an edge waveguide is initially excited. 

Also, from lower rows in Figure \ref{fig.semiinfinite1} and Figure \ref{fig.finite1}, 
 it can be appreciated that (when present) reflections at the edge waveguides can be tailored by adjusting the non-Hermitian parameter $\gamma$. Moreover, it is seen that in such boundaries or edge waveguides the non-Hermitian skin effect might be present, for instance as shown in the lower middle panels of Figure \ref{fig.semiinfinite1} and Figure \ref{fig.finite1}, and the lower right panel of Figure \ref{fig.finite1}, due to the non-unitary transformation (\ref{phipsi}). In turn, the skin effect has interesting potential applications, for instance in optical devices (see Ref. \cite{Weidemann2020}).

In addition, along this manuscript was considered that the non-unitary transformation (\ref{phipsi}) is applied to the initial state $|\psi(0)\rangle$ of the initial Hermitian system. However, and as mentioned before, such transformation can be applied at an arbitrary value of the  propagation distance $z$. The inverse transformation (\ref{psiphi}) can be performed at any $z$ as well, taking us back to the original Hermitian system. Therefore, in principle it is possible to alternate intervals of Hermitian and non-Hermitian propagation of the electromagnetic field at will, by the action of the $e^{\pm\gamma \hat n}$ operators (see Figure \ref{fig.transport}), which in turn might be useful in the management and encoding of optical information, for example.

Finally, the (rather simple) transformation (\ref{gphipsi}) gives the solution $|\phi(z)\rangle$ of the non-reciprocal system (\ref{nHermitian})-(\ref{htilde}), for more general $k_1$ and $k_2$ than those considered in the present work (those leading to the Hatano-Nelson type problem), in terms of the solution $|\psi(z)\rangle$ of the corresponding reciprocal (isotropic) Hermitian system (\ref{quantum})-(\ref{Hamilt}), for $\alpha=\sqrt{k_1 k_2}$. In turn, such anisotropy can be achieved, for instance, by considering that the cross section of the waveguides in the array either increase or decrease with the site number $n$, according to the definition of the coupling coefficients (see Eq. (13.2-15) in Ref. \cite{Yariv} and Eq. (13.3-9) for the case of two waveguides). A deeper insight on this matter will be reported elsewhere.

\end{document}